# POLARIS: A framework to guide the development of Trustworthy AI systems

Maria Teresa Baldassarre
University of Bari "A. Moro"
Bari, Italy
mariateresa.baldassarre@uniba.it

Domenico Gigante
Ser&Practices Srl
Bari, Italy
d.gigante@serandp.com

Marcos Kalinowski
Pontifical Catholic University of Rio de Janeiro (PUC-Rio)
Rio de Janeiro, Brazil
kalinowski@inf.puc-rio.br

Azzurra Ragone
University of Bari "A. Moro"
Bari, Italy
azzurra.ragone@uniba.it

## ABSTRACT

In the ever-expanding landscape of Artificial Intelligence (AI), where innovation thrives and new products and services are continuously being delivered, ensuring that AI systems are designed and developed responsibly throughout their entire lifecycle is crucial. To this end, several AI ethics principles and guidelines have been issued to which AI systems should conform. Nevertheless, relying solely on high-level AI ethics principles is far from sufficient to ensure the responsible engineering of AI systems. In this field, AI professionals often navigate by sight. Indeed, while recommendations promoting Trustworthy AI (TAI) exist, these are often high-level statements that are difficult to translate into concrete implementation strategies. Currently, there is a significant gap between high-level AI ethics principles and low-level concrete practices for AI professionals. To address this challenge, our work presents an experience report where we develop a novel holistic framework for Trustworthy AI — designed to bridge the gap between theory and practice — and report insights from its application in an industrial case study. The framework is built on the result of a systematic review of the state of the practice and a survey and think-aloud interviews with 34 AI practitioners. The framework, unlike most of those already in the literature, is designed to provide actionable guidelines and tools to support different types of stakeholders throughout the entire Software Development Life Cycle (SDLC). Our goal is to empower AI professionals to confidently navigate the ethical dimensions of TAI through practical insights, ensuring that the vast potential of AI is exploited responsibly for the benefit of society as a whole.

## 1 INTRODUCTION

In the dynamic realm of Artificial Intelligence (AI), marked by ceaseless innovation and rapid advancements, the ethical, societal, and operational implications of AI technologies have shifted to the forefront of discussions. As AI systems become deeply integrated into our daily lives, from health-care [1] to finance [2], and influence critical decision-making processes, the responsible development and deployment of AI has transitioned from an academic discourse to an imperative in real-world applications.

In this complicated context, the concept of Trustworthy Artificial Intelligence (TAI) has emerged. We relate to the following definition: *"Trustworthy AI has three components, which should be met throughout the system's entire life cycle: (1) it should be lawful, complying with all applicable laws and regulations (2) it should be ethical, ensuring adherence to ethical principles and values and (3) it should be robust, both from a technical and social perspective since, even with good intentions, AI systems can cause unintentional harm. Each component in itself is necessary but not sufficient for the achievement of Trustworthy AI"* [3]. These risks are even more pronounced with the recent advent of Generative AI — *e.g.* ChatGPT — and how this impacts on various societal aspects [4].

---

*Authors are listed in alphabetical order.

The significance of actively considering Trustworthy AI cannot be overstated. While AI carries the transformative potential to revolutionize industries and drive societal progress, it equally bears inherent risks. AI systems, if left unchecked, can perpetuate biases, threaten individual privacy, and undermine cybersecurity, potentially resulting in unintended societal harm. They have the power to significantly influence decisions affecting individuals and communities, and in the wrong hands, can lead to catastrophic consequences. Trustworthy AI provides a safeguard against such risks, ensuring that as AI continues its evolution, it does so in an ethical, transparent, and equitable manner while adhering to stringent standards for privacy and fairness.

Several public and private organizations have tried to address TAI by developing different kinds of resources: ethical requirements [5], principles [6], guidelines [7], best practices [3], tools [8], and frameworks [9]. However, navigating the intricacies of TAI has become increasingly complex due to what we might refer to as "*principle proliferation*" [10]. This phenomenon encompasses the multitude of ethical principles that have been devised, each one providing a specific definition, but also contributing to a landscape that can overwhelm AI practitioners.

In response to the challenges posed by principle proliferation, our research follows the work of Jobin et al. [10] and focuses on four foundational pillars of TAI: Privacy, Security, Fairness, and Explainability. These pillars have been condensed in a practical, focused, and adaptable framework, called POLARIS. Its aim is to guide AI practitioners and stakeholders in their quest to ensure the effective trustworthy development of AI-enabled systems across the entire software Development LifeCycle (SDLC).

In this experience report paper, we explain the genesis of the framework that followed an inductive process: first, we conducted a *systematic literature review* of the state of the practice to understand the TAI frameworks that have been proposed by industry, academia, and other institutions. As a result, we highlighted several missing points and an evident gap between theory (high-level principles) and practice (concrete implementation strategy). Then, we distributed a *survey* to AI practitioners, to understand their needs, desires and the challenges they encounter when trying to build trustworthy AI applications. We build on this collected knowledge to design the POLARIS framework and then apply it to an industrial case. The main contributions of this paper can be summarized as follows:

- A review of the state of the practice and identification of practitioner needs to understand existing practices, challenges, and what practitioners currently lack in developing Trustworthy AI applications.
- The proposal of a novel framework (POLARIS) that systematizes and organizes the knowledge found in different sources. The objective of POLARIS is to make this knowledge easily accessible to AI practitioners and to provide them with actionable guidelines that can be applied in every phase of the SDLC.
- A first validation of the framework through an industrial case study.

The paper is organized as follows: Section 2 provides some useful background definitions about Trustworthy AI principles, while Section 3 outlines the results of the systematic review and the findings from the survey and interviews with AI professionals. Section 4 describes the POLARIS framework, its components and how to use it. Section 5 provides the preliminary results obtained after applying the framework to an industrial project, as well as the changes applied and the lessons learnt. Section 6 addresses the limitations of our framework and finally conclusions are drawn in Section 7.

## 2 BACKGROUND

This section encompasses some preliminary definitions needed to understand the context of this work.

### 2.1 AI Principles proliferation

National and international entities have established specialized expert committees within the domain of Artificial Intelligence (AI) to proactively address the manifold risks associated with AI development. These committees are often vested with the responsibility of crafting policy documents and recommendations. Notable examples of such organizations include the High-Level Expert Group on Artificial Intelligence, an initiative spearheaded by the European Commission [11], the UNESCO Ad Hoc Expert Group (AHEG) assigned the task of formulating the Recommendation on the Ethics of Artificial Intelligence [12], the Advisory Council on the Ethical Use of Artificial Intelligence and Data in Singapore [13], the NASA Artificial Intelligence Group [14], and the UK AI Council [15], among others.

These committees bear the crucial role of generating comprehensive reports and guidelines about Trustworthy AI (TAI). A parallel endeavour is observable within the commercial landscape, particularly among enterprises heavily reliant on AI technologies. Corporations such as Sony[1] and Meta[2] have made their AI policies and principles publicly accessible. Concurrently, professional organizations and non-profit entities, such as UNI Global Union[3] and the Internet Society[4], have issued statements and recommendations.

The substantial efforts of this diverse spectrum of stakeholders in crafting TAI principles and policies not only underscore the imperative for ethical guidance but also exemplify their vested interest in shaping AI ethics to align with their specific objectives [16]. It is noteworthy that the private sector's engagement in the gap of AI ethics has undergone a thorough check, with contentions suggesting that high-level soft policies may be employed to either transform a social issue into a purely technical one [16] or to potentially circum- vent regulatory measures [10, 17].

---
[1]https://www.sony.com/en/SonyInfo/sony_ai/responsible_ai.html
[2]https://ai.facebook.com/blog/facebooks-five-pillars-of-responsible-ai/
[3]http://www.thefutureworldofwork.org/media/35420/uni_ethical_ai.pdf
[4]https://www.internetsociety.org/resources/doc/2017/artificial-intelligence-and-machine-learning-policy-paper/

Nevertheless, a set of research endeavours has brought attention to the divergent nature of these proposals, giving rise to a complex challenge often referred to as *"principle proliferation"* [18]. Consequently, efforts have been undertaken to address this challenge. For instance, Jobin et al. [18] conducted a comprehensive study, that culminated in the identification of a global convergence around five ethical principles: *transparency*, *justice and fairness*, *non-maleficence*, *responsibility*, and *privacy*. Jobin et al. [18] observed that, while no single document they reviewed encompassed all of these ethical principles, these five principles were mentioned in over half of the sources examined. Furthermore, their detailed thematic analysis unveiled significant semantic and conceptual variations in the interpretation of these principles and the specific recommendations or areas of concern derived from each one.

## 2.2 Trustworthy AI principles: definitions

As set out in Section 2.1, a notable degree of ambiguity and subtlety exists in demarcating the principles that predominantly characterize Trustworthy AI (TAI). Notably, TAI is sometimes used interchangeably with Responsible or Ethical AI. In our investigation, we confront the challenge of *principle proliferation* by choosing to focus on a specific subset that characterizes TAI. Specifically, we concentrate on the most recurrent four principles identified by Jobin et al. [10], while opting to exclude the principle of *responsibility* due to its infrequent occurrence and lack of a clear, universally accepted definition.

Furthermore, in this work we have decided to adopt the definitions put forth by the High-Level Expert Group on Artificial Intelligence[5] — an entity established by the European Commission [5] — explained in their "*Ethics guidelines for trustworthy AI*" [3].

Given these premises, for each of the four selected principles among those claimed by Jobin et al. [10], we map the formal definition delineated in the guidelines that we adopt in our study [3]. Mapping has been carried out based on the contents of the definitions and not merely on the nomenclatures as, in most cases, they differ.

Furthermore, for the sake of simplicity, we have labelled each TAI principle as follows: Fairness, Explainability, Security, and Privacy. We will use these labels throughout the paper.

Explainability, also referred to as *transparency* [3, 10].

«[...] Explainability concerns the ability to explain both the technical processes of an AI system and the related human decisions (e.g. application areas of a system). Technical explainability requires that the decisions made by an AI system can be understood and traced by human beings. [...]. [The] explanation should be timely and adapted to the expertise of the stakeholder concerned (e.g. layperson, regulator, or researcher). [...]» [3]

This principle includes and can be directly associated with system requirements such as *traceability* and *explainability*.

Fairness, referred to as *Diversity, non-discrimination and Fairness* [3] or *Justice and Fairness* [10].

«In order to achieve Trustworthy AI, [one] must enable inclusion and diversity throughout the entire AI system's life cycle. [...] this also entails ensuring equal access through inclusive design processes as well as equal treatment. [...] Bias [derives from] data sets used by AI systems (both for training and operation) [because these] may suffer from the inclusion of inadvertent historic bias, incompleteness, and bad governance models. The continuation of such biases could lead to unintended (in)direct prejudice and discrimination against certain groups or people, potentially exacerbating prejudice and marginalization. [...]» [3]

This principle can be directly mapped with system requirements such as *avoidance of unfair bias*, *accessibility and universal design*, and *include stakeholder participation*.

Security, identified as *Technical robustness and safety* [3], and as *Non-maleficence* [10].

« [...] Technical robustness requires that AI systems are developed with a preventative approach to risks and in a manner such that they reliably behave as intended while minimizing unintentional and unexpected harm, and preventing unacceptable harm. This should also apply to potential changes in their operating environment or the presence of other agents (human and artificial) that may interact with the system in an adversarial manner. In addition, the physical and mental integrity of humans should be ensured.» [3]

This principle can be directly mapped with system requirements such as *resilience to attack and security*, *fallback plan*, *general safety*, *accuracy*, *reliability*, and *reproducibility*.

Privacy [10], also referred to as *Privacy and data governance* [3].

« [...] Privacy [is] a fundamental right particularly affected by AI systems. Prevention of harm to privacy also necessitates adequate data governance that covers the quality and integrity of the data used, its relevance in light of the domain in which the AI systems will be deployed, its access protocols and the capability to process data in a manner that protects privacy.» [3]

This principle can be directly mapped with system requirements such as *including respect for privacy*, *quality and integrity of data* and *quality and integrity of access to data*.

## 3 STATE OF THE PRACTICE

This experience report builds on previous research [19], in which we conducted a comprehensive study of the state of the practice of existing TAI frameworks. More precisely, we investigated (i) the extent to which the analyzed frameworks addressed the principles mentioned in Section 2.1 and (ii) whether and to what extent these frameworks covered the stages of the Software Development Life Cycle (SDLC). Next, we carried out a comparative analysis among the identified frameworks with respect to characteristics such as best practices, guidelines, and tools, in order to assess if and how big of a gap there is between the proposed high-level AI ethics principles and low-level operational practices for practitioners.

---
[5]https://digital-strategy.ec.europa.eu/en/policies/expert-group-ai

In our previous work [19] we analyzed 138 frameworks, both from white-literature and grey-literature sources. The main findings are:

1) Most of the frameworks are proposed by *No-profit Organisations, Public Entities or Human Communities* (50.7%); followed by private Companies (31.9%) and then Universities (17.4%).

2) Most of the frameworks provide a set of TAI principles/values (46.1%); others include actionable Guidelines (29.6%), but very few also provide Tools (9.2%).

3) In the majority of cases, *the frameworks address all four TAI principles* (45.1%) even if there are frameworks that cover only one (15.5%) or two (15.5%) principles;

4) More than half of the frameworks (55.2%), provide support only for the *Requirements Elicitation* phase. While all the SDLC phases are covered only in 5.7% of the frameworks;

5) In more than 80% of the cases there is *no tool included in the framework* and when it is present, it is directed to *Non-technical* stakeholders (i.e. stakeholders who work in the first two phases of the SDLC — e.g. commercial agents, functional analysts, architecture designers, etc).

In summary, our analysis confirmed that most of the existing frameworks include high-level best practices, checklists, or self-assessment questions, most suitable for non-technical stakeholders, limitedly able to address technical stakeholder needs and close the gap between high-level principles definition and practical recommendations for AI practitioners covering all the SDLC phases. Moreover, the findings [19] have guided our subsequent research aimed at investigating the needs of AI practitioners, and their current practices and issues encountered in the design and implementation of trustworthy AI systems.

## 3.1 Identification of Practitioner needs

As a premise for the design and proposal of our framework, we conducted an exploratory survey [20] to collect practitioner insights and needs with respect to TAI principles. We used convenience sampling and recruited practitioners from companies in our network of collaborations. All participants were practitioners with experience in developing AI-enabled systems who had addressed, to some extent, TAI in their projects. We contacted a total of 45 professionals of which, 34 completed the survey. These participants represented a diverse spectrum, ranging from small-medium companies, (55.9%) of various dimensions, to large companies, with more than 1000 employees (44.1%).

Apart from the demographics, the survey is organized into three main parts, each focusing on the collection of specific data pertaining: a) existing practices, b) identification of challenges, c) discovery of unmet needs. In the following, for each part, we provide a brief explanation and highlight the main results.

*A) Exploration of Existing Practices*. The first part of the survey investigated the operational procedures and methodologies employed by practitioners in the context of implementing TAI. This exploration aimed to provide a detailed insight into the real-world practices and strategies adopted. Results. First, we observed that the TAI principle most frequently addressed by participants is *Privacy* (58.8%), and most of the participants address at least one TAI principle during *Design* (64.7%) and *Development* (47.1%) SDLC phases. On the contrary, very few participants declared to address at least one TAI principle during the *Test* (29.4%) and *Deploy* (20.6%) phases. This may highlight the need for more support, in terms of tools and guidelines in the last phases of the SDLC. Moreover, when participants faced issues related to TAI, in half of the cases they did not even try to address or solve them (50%). Probably because they did not know how to or they simply considered them not worth solving. This is a point that deserves more investigation. Only 35% of the participants declared to have directly addressed TAI issues, while a small percentage stated that the issue resolution was demanded to a third party (15%).

*B) Identification of Challenges*. This section of the survey explored the challenges and obstacles encountered by professionals while trying to integrate TAI into their systems. By identifying these issues, we aimed to shed light on critical areas where AI professionals may need more support. Results. In cases where the respondents tried to address/fix TAI issues, we found that the most voted impediments are: (i) "*the issue solution required too much time to be implemented*" (58.3%) and (ii) "*the issue solution was likely to decrease the performance of the system (e.g., decreasing accuracy)*" (50%). On the other hand, none of the participants answered: "*no one had idea on how to solve the issue*", which is a positive result since it indicates that practitioners are conscious of untrustworthiness problems and are able to hypothesize solutions. Among the comments, one participant mentioned "*[scarce] data availability*" as an impediment.

*C) Discovery of Unmet Needs*. The third part of the survey reveals the presence of unaddressed needs within the practical landscape of TAI. Specifically, we uncovered a range of requirements that have so far received limited attention within the existing literature.

Results. Regarding the prevention of trustworthiness issues in AI, the participants rated as the most valuable tool ableto "*[...] generate an explanation of a model after its creation [...]*" (with 82% of positive answers) while they rated as least useful (i) a tool to help "*deciding how much data you need for particular subgroups/subpopulations*" and (ii) a tool to "*generate possible adversarial/malicious data points to test to use in testing the system*" (both with 19% of negative answers).

On the other side, to address untrustworthiness in AI, the participants rated as the most valuable (i) "*best practices that can actively guide your team through the model's SDLC* "(92% positive answers), (ii) a tool able to "*[...] help [...]*

*monitoring the AI model after its release to the public*" (91% positive answers), and "*a knowledge book in which are mapped trustworthiness problems and [...] solutions*" (70% positive answers). On the other hand, they rated as least useful (i) a tool "*[...] to help your team doing an ex-post TAI audit*" (18% of negative answers) and (ii) a tool able to *[...] help your team deciding which AI model best respects the TAI principles [...]*" (17% of negative answers).

Overall, these results confirmed the findings of our previous work [19]: a significant majority of the respondents expressed the need for *comprehensive knowledge bases* and *pragmatic guidelines* offering insights and recommendations for the seamless implementation of trustworthy AI system throughout the entire SDLC. Furthermore, they also highlighted the lack of *tools* supporting them in the last stages of SDLC.

## 4 THE POLARIS FRAMEWORK

In response to the challenges and issues highlighted by the research results, with the intent to fill the gap between theory and practice and to address stakeholder needs and shortcomings (Section 3), we have developed a framework: POLARIS.

Indeed, POLARIS has been designed to provide actionable guidelines and tools in order to support stakeholders in addressing TAI principles throughout the entire Software Development Life Cycle (SDLC). POLARIS provides a significant amount of information, organized and linked into a comprehensive knowledge base that is designed to be expandable, with the possibility to easily add new knowledge.

In Section 4.1 we explain how we built the POLARIS knowledge base while in Section 4.2 we describe how to navigate it.

### 4.1 Defining POLARIS Knowledge Base

In this section, we describe how we assembled the POLARIS Knowledge Base and the selection process used to choose the different knowledge sources representing the foundation of POLARIS.

We started from the frameworks analyzed in [19], we complemented our analysis with the results obtained from the survey, and then we identified among the existing knowledge sources (i.e. frameworks) those that met both of the following criteria:

(1) Have actionable guidelines (and not only a simple high-level principles list)
(2) Address all SDLC phases.

Regarding the last criterion, since SDLC phases do not always map with the activities required to develop an AI-enabled system, we have integrated each SDLC phase with AI-enabled activities established by Zhengxin et al. [21]. Table 1 shows how each SDLC phase has been integrated with AI-enabled system development activities.

After this first selection phase, we identified only three knowledge sources that meet both criteria (1) and (2). Then, we mapped each identified knowledge source to the corresponding TAI principle. For Explainability we selected Jin et al. - EUCA: the Explainable AI Framework [22], for Fairness we chose Amsterdam Intelligence - The Fairness Handbook [23]. For both Privacy and Security, we selected ENISA - Securing Machine Learning Algorithms [24].

Table 1: Activities integrated into the SDLC phases to support AI-enabled systems development.

| SDLC phase | AI activity |
| --- | --- |
| Requirements Elicitation | Model Requirement |
| Design | Data Collection, Data Preparation |
| Development | Feature Engineering, Model Training |
| Test | Model Evaluation |
| Deployment | Model Deployment |
| Monitoring | Model Monitoring |

Then, we refined this first selection by adding more knowledge sources that could complement the information provided by the primary ones initially selected. We started by selecting the frameworks that met *at least one* of the following criteria:

(1) Have actionable guidelines (and not only a simple high-level principles list)
(2) Address all SDLC phases.

We retrieved 10 additional knowledge sources that met at least one of the previous criteria. The table with all the 10 knowledge sources identified can be found in the online appendix [29]. Then, we performed a comparative analysis between each primary knowledge source already selected ([22], [23], [24]) and the new ones retrieved in this second iteration.

The results of the comparative analysis brought us to select four additional knowledge sources that could complement and expand the information provided by the first ones selected. The additional frameworks selected were ICO's "*Guidance on AI and data protection*" [27], Tensorflow's "*Responsible AI in your ML workflow*" [25], the guidelines in Microsoft's "*Threat Modeling AI/ML Systems and Dependencies*" [28] and CSIRO's "*Responsible AI Pattern Catalogue*" [26]. Therefore, we used these additional knowledge sources to further extend the information provided by the primary ones. We ended up selecting 7 knowledge sources. Table 2 shows the mapping between each TAI principle and the corresponding knowledge sources covering that principle.

### 4.2 Navigating POLARIS Knowledge Base

Having defined the POLARIS knowledge base, in this section we focus on how to navigate it. The goal of the proposed framework is to support stakeholders throughout the SDLC by suggesting concrete implementation strategies able to support and guide them in the development of TAI applications.

When applying the framework, the users will ultimately receive an *Action* to implement, that is, an actionable guideline that a stakeholder should consider and, if possible, implement while developing the AI-enabled software system to ensure compliance with the four TAI principles. The user can also choose to filter and apply only a subset of the suggested guidelines.

Table 2: The identified *best knowledge source* for each TAI principle.

| TAI Principle | Knowledge Source |
| --- | --- |
| Explainability | Jin et al. - EUCA: the Explainable AI Framework [22] |
| | Tensorflow - Responsible AI in your ML workflow [25] |
| | CSIRO - Responsible AI Pattern Catalogue [26] |
| Fairness | Amsterdam Intelligence - The Fairness Handbook [23] |
| Security | ENISA - Securing Machine Learning Algorithms [24] |
| | ICO - Guidance on AI and data protection [27] |
| | Tensorflow - Responsible AI in your ML workflow [25] |
| | Microsoft - Threat Modeling AI/ML Systems and Dependencies [28] |
| | CSIRO - Responsible AI Pattern Catalogue [26] |
| Privacy | ENISA - Securing Machine Learning Algorithms [24] |
| | ICO - Guidance on AI and data protection [27] |
| | Tensorflow - Responsible AI in your ML workflow [25] |
| | Microsoft - Threat Modeling AI/ML Systems and Dependencies [28] |
| | CSIRO - Responsible AI Pattern Catalogue [26] |

As of now, the first version of POLARIS has been structured as a filterable Excel array of sheets. There are four main knowledge components, one per each principle: (i) *Privacy*; (ii) *Security*; (iii) *Fairness* and (iv) *Explainability*.

In proposing the structure of each Excel sheet, we were inspired by the ENISA framework [24] and then customized it according to our needs.

The two sheets that contain the knowledge for "Privacy" and "Security" are composed of the following six columns (Fig. 1 shows an excerpt of the security component).

(1) SDLC Phase. The SDLC phase that the *Action* column applies to.

(2) Threat. Contains the list of threats, i.e. possible attacks that can be conducted against an AI-enabled system. Examples are *Evasion* and *Poisoning* attacks.

(3) Sub-Threat. In some specific cases, a threat can have a specific declination in a sub-characteristic. For example, the *Poisoning* attack can be declined in *Targeted Data Poisoning* and *Indiscriminate Data Poisoning*.

(4) Description. A textual description of the (Sub)Threat, which helps the stakeholder obtain coarse-grained details about the threat and understand the attacker's objective.

(5) Vulnerability (consequence). This is the immediate consequence of having a model vulnerable to a specific threat.

(6) Action. The corresponding action, or decision, that should be adopted to address a specific threat, based on the SDLC phase and threat selected, keeping in mind the vulnerability.

For example, a developer in the *Design* SDLC phase who is trying to address the vulnerabilities associated with *Poisoning* threat, may consult POLARIS and access the *Security* Excel sheet, select the *Poisoning* threat — and corresponding sub-threat, i.e. *Label modification* —, and obtain a description of the vulnerability associated to the (sub)threat and the action to take in order to mitigate the vulnerability, i.e. *ensure that reliable sources are used* (Fig. 1).

The sheet that contains the knowledge for "Fairness" is composed of 5 columns, all of the above, except for *Vulnerability (consequence)* column which has been removed as the concept of vulnerability in the context of fairness does not apply.

The sheet containing the knowledge for "Explainability" has a different set of columns (see Fig. 2), because there are no real threats associated with the lack of explainability. However, having a system that is not explainable, will lead users to use it with some reluctance because of its opacity in making decisions, as it is not possible to derive any clear logical relationship between the internal configuration and their external behaviour, except for a few specific cases (e.g. decision trees) [30].

For Explainability, the columns are the following:

(1) SDLC Phase. The SDLC phase the *Action* column relates to.

(2) Data Type. The type of data used by the AI algorithm for which the action/guideline applies. Examples are *Tabular* data or *Image*. When the action applies to all algorithms, regardless of the type of data, the tag *General* is used.

(3) Local/Global Explanation. This column describes the type of explanation that can be obtained by implementing the action. At the moment, the possible values are *Global* and *Local* [31].

(4) Explanation Goal. This is the goal that can be achieved if the action/guideline gets implemented. Examples are: *to validate the algorithm outcome* and *to reveal bias*.

(5) Action. The corresponding action, or decision, that should be taken to reach the selected explanation goal. We point out that for each <data, explanation type> pair there is at least a corresponding row in the framework.

For example, a user who is in the *Requirement Elicitation* SDLC phase and needs to enquire on all the possible explanation approaches to explain the output of an algorithm, could access the *Explainability* Excel sheet and select the *General*

POLARIS: A framework to guide the development of Trustworthy AI systemsFigure 1: Excerpt of the Security component navigation of the POLARIS framework.

| SDLC Phase | Threat | Sub-Threat | Description | Vulnerability (consequence) | Action |
|---|---|---|---|---|---|
| Design | Poisoning | Label modification | An attack in which the attacker corrupts the labels of training data. This sub-threat is specific to Supervised Learning. | Use of unreliable sources to label data | **(Technical) Ensure reliable sources are used**: ML is a field in which the use of open-source elements is widespread (e.g., data for training, including labeled ones, models). The trust level of the different sources used should be assessed to prevent using compromise ones. For example: the project wants to use labeled images from a public library. Are the contributors sufficiently trusted to have confidence in the contained images or the quality of their labelling? |
| Development | Failure or malfunction of ML application | Denial of service due to inconsistent data or a sponge example | ML algorithms usually consider input data in a defined format to make their predictions. Thus, a denial of service could be caused by input data whose format is inappropriate. It may also happen that a malicious user of the model constructs an input data (a sponge example) specifically designed to increase the computation time of the model and thus potentially cause a denial of service. | Use of uncontrolled data | **(Technical) Control all data used by the ML model** Data must be checked to ensure they will suit the model and limit the ingestion of malicious data: <br>- Evaluate the trust level of the sources to check it's appropriate in the context of the application <br>- Protect their integrity along the whole data supply chain <br>- Their format and consistence are verified <br>- Their content is checked for anomalies, automatically or manually (e.g. selective human control) <br>- In the case of labeled data, the issuer of the label is trusted. |

Figure 2: Excerpt of the Explainability component navigation of the POLARIS framework

| SDLC Phase | Data type | Local/Global Explanation | Explanation Goal | Action |
|---|---|---|---|---|
| RE | General | Both | Start considering Explainability from Req. Elicitation | Elicit also explainability requirements; examples are: <br>- Unexpected Prediction: Disagreement with AI: declare the required behaviour in case the AI prediction is unexpected, and/or users disagree with AI's prediction <br>- Expected prediction: declare the required behaviour in case AI's prediction aligns with users' expectations <br>- Differentiate similar instances: due to the consequences of wrong decisions, users sometimes need to discern similar instances or outcomes. For example, a doctor differentiates whether the diagnosis is a benign or malignant tumor <br>- Learn from AI: users need to gain knowledge, improve their problem-solving skills, and discover new knowledge <br>- Improve the predicted outcome: users seek causal factors to control and improve the predicted outcome <br>- Communicate with stakeholders: many critical decision-making processes involve multiple stakeholders, and users need to discuss the decision with them <br>- Generate reports: users need to utilize the explanations to perform particular tasks such as report production. For example, a radiologist generates a medical report on a patient's X-ray image |
| Design | General | Both | Have a clear idea about the desired explanation form | Consider the design of the explanation design: how the final UI should be composed and how to present the information |

data type and retrieve a set of actions that pertain explainability requirements i.e. *elicit explainability requirements* (Fig. 2).

When navigating POLARIS, each stakeholder can use different filters and subfilters, based on specific needs, as for instance: *Knowledge Component* (i.e. TAI principle) to address, *Threat* (or *Sub-Threat*), *Vulnerability*, *SDLC Phase*, *Data type*, and *Local/Global Explanation*.

One of the most significant filters is *SDLC phase*, which makes POLARIS flexible and allows stakeholders to use it either on ongoing/closed projects — where it is possible to address, for example, only the deployment or monitoring phase — or at the early stage of a project, since in the latter case it can cover all SDLC phases.

## 5 POLARIS FRAMEWORK APPLICATION

With the aim of validating POLARIS framework and investigating key points for improvement, we have applied the framework in an industrial project and collected feedback from practitioners through a think-aloud session as they applied the framework during the development of an AI-enabled application. Section 5.1 gives more details about the industrial case, while Section 5.2 describes the improvements made after having collected feedback. Finally, Section 5.3 describes the lessons learnt from applying the framework to a first industrial real case.

### 5.1 Industrial Case setting

We applied POLARIS during the development of a real-world software application on behalf of a SME. In accordance with the guidelines provided by Runeson and Höst [32], the research question (RQ) we aimed to address questioned: *Is POLARIS helpful in supporting different stakeholders dealing with Trustworthy AI challenges in all the SDLC phases?* With this RQ in mind, we focused on how and to what extent POLARIS helps stakeholders. Furthermore, we take the feedback received to improve the framework itself. During the development of the application, developers used POLARIS and were, at the same time, involved in think-aloud sessions to collect feedback.

The case selection involved an AI-enabled software application developed by an IT SME that mainly provides services in various domains such as cybersecurity, AI, and Software Engineering. The IT team of the company is composed of 7 seniors, 3 mid-experienced, and 15 juniors.

The application was commissioned to the company by a public entity that was particularly committed to AI ethics and trustworthiness issues. The software application had to perform image recognition and object detection tasks. Moreover, since the application is used in a very sensitive and critical domain, the system could not make autonomous decisions but had to only provide suggestions to the human operator — functioning as a decision support system. One of the requirements was that each suggestion must be supported by a graphical, self-explainable and human-readable description. This application required dealing with *computer vision*, specifically *image analysis*, so includes methods for acquiring, processing, analyzing, and understanding digital images [33].

The IT teams chose to use a deep-learning model to perform these tasks. This meant that several images were necessary to conduct a proper training phase and achieve high-level performances. Since training data was not available, except for only a few samples, the developers artificially created a dataset starting from the images available in four open-source datasets.

Due to the sensitivity of the domain, the entire system was deployed on an on-premise infrastructure.

While designing and developing the application, developers applied the POLARIS framework components during all phases of the SDLC.

The data collection (3) was conducted by means of the concurrent thinking-aloud method [34, 35]: we first explained to a software development team how POLARIS was structured and should be integrated into their work; then, we participated in their daily routines (on days agreed upon with them) and asked them to think aloud as they used POLARIS during all moments of the SDLC. The team was composed of 1 Business Analyst (BA) — who tested POLARIS Fairness and Explainability components during the entire Requirement Elicitation phase — 1 Senior Software and Data Engineer (SDE) — who tested Fairness and Explainability during the Design and Development phases — 1 Junior Developer (DEV) — who tested Fairness and Explainability during the entire Development phase — 1 Cybersecurity Specialist & DevOps (CS) — who tested Privacy and Security during the entire SDLC — Quality Assurance Engineer (QA) — who tested Fairness and Explainability during the entire Test phase — and 1 IT engineer (IT) — who tested Fairness and Explainability during the Deploy and Monitoring phases. During sessions, we took notes. In particular, the researcher reminded participants to think aloud and solicited comments through questions to motivate feedback.

The data analysis (4) was conducted by means of qualitative methods. Once the software system was delivered to the final customer, we analyzed all the collected feedback notes to conclude to what extent POLARIS supports managing TAI principles and establish areas of improvement. The main outcomes of the study along with the improvements applied to the framework, are reported in Section 5.2.

As recommended by Creswell [36], to ensure the validity of the findings, we followed the *member check* strategy and asked all team members involved in the study to validate our findings and provide feedback on the sessions.

## 5.2 POLARIS evolution

Once the study was completed, we used the results collected from the feedback sessions to implement a set of changes intended to improve POLARIS.

In this section, we outline the most important findings that have led to changes made to POLARIS and explain the rationale behind each one. The complete list of changes can be found in the online appendix [29].

Adding filtering capabilities. The initial POLARIS version did not have any filtering options. It was a single Excel file, thus the users must vertically scroll the entire sheet to find the relevant information for a specific component or (sub-)threat. Moreover, there was a column for each SDLC phase.

However, all team members reported that POLARIS was very cumbersome and difficult to use with these settings. So, the first improvements were related to usability: we created a single file for each TAI principle and inserted a dropdown list allowing to to choose the SDLC phase of interest. Other filtering options have also been applied. For space reasons, we have not detailed them.

Revise the mapping between SDLC phase and guidelines. SDE, DEV, CS, QA and IT reported that it was necessary to review the SDLC phase with which some guidelines were mapped. In particular, there were guidelines (e.g., "*Implement processes to maintain security levels of ML components over time*" and "*Include ML applications into detection and response to security incident processes*") that were too abstract to be mapped to practical SDLC phases (e.g., *Development* and *Test*).

Conversely, some guidelines were mapped only with the later phases of the SDLC, while our testers highlighted the need for help early on in the process. For example, the guideline "*Define and monitor indicators for proper functioning of the model*", according to ENISA, was mapped only with *Monitoring* phase; we also added the *Design* phase because it is important to start defining the *Key Performance Indicators* (KPIs) even before starting the development of the model itself: once the model is developed, missing metrics can be discovered but modifying the code may now cost too much time and/or money.

Remove TAI patterns. Initially, we included all TAI patterns provided in Data61-CSIRO's "*Responsible AI Pattern Catalogue*" [26] in POLARIS, because we assumed every pattern could help developers to incorporate TAI in the development phase. Feedback from SDE, DEV, and CS, when applying these patterns, pointed out that some patterns were not really applicable: for example, the pattern "*TAI user story*" gets implemented simply by using POLARIS itself. We revised and removed the patterns that were not considered relevant or useful.

Remove *governance-oriented* guidelines. All team members pointed out that there were guidelines that cannot really be implemented through the SDLC phases, but require an organizational posture change, with new aspects to be considered in their internal policies. This happened, for example, for the guideline "*Integrate ML specificities to awareness strategy and ensure all ML stakeholders are receiving it*". So, we decided to remove such guidelines from POLARIS because the framework is not meant to provide suggestions at the organization's policy level.

Remove equivalents guidelines. Since we included guidelines from different knowledge sources, we initially missed the fact that diverse guidelines may address the same (sub-)threat and provide the same mitigation but with different words. After SDE, DEV, CS, QA, and IT reported this situation, we investigated and removed these redundancies. For example, Microsoft's guideline "*Input validation, both sanitization and integrity checking*" was deleted because almost identical to the ENISA's "*Control all data used by the ML model*".

Highlight the fact that some guidelines provide different alternative mitigations for the same threat. Especially in the Development and Test phases of the SDLC, there are a lot of alternative tools to automatically address a specific threat. SDE, DEV, IT, and QA pointed out that a less familiar stakeholder may miss this nuance and implement all suggestions, discovering only in the end that all of them led to the same result. So, we decided to make clearer the cases where the user can simply choose one alternative instead of applying the entire array of mitigations.

Contextualize abstract guidelines. Analyzing the different knowledge sources, SDE and CS found that some guidelines sound too abstract to be brought back to something concretely implementable. We reviewed these cases, analyzing the literature to find the best actionable mitigations, and integrated them into POLARIS. For example, this process was conducted for the ENISA's guideline "*Choose and define a more resilient model design*": here we added more specific and concrete examples by integrating the knowledge provided in [37].

Merge similar guidelines. In the Security component — where we included the guidelines proposed by both Microsoft [28] and ENISA [24] — CS found that most of the guidelines proposed by Microsoft addressed the same threat already addressed by ENISA; in some cases, Microsoft proposed one or few additional actions compared to ENISA (or vice-versa). To address this issue, we decided to take the ENISA's guideline — because was the most comprehensive — and merge into it the missing details from Microsoft's guideline. For example, this happened for the "*Lack of training based on adversarial attacks*" vulnerability, linked to *Evasion* threat.

Add missing tools. While using the framework, DEV, CS, QA, and IT reported to us that they found some guidelines that were mainly descriptive and did not provide any tools useful for implementing those guidelines. Since in many cases, they knew there was a tool to support mitigation strategies, we added all the tools they suggested. For example, we added ENISA's online tool for the security of personal data processing[6] for conducting DPIA and support the guideline "*ICO 1.1 - Conduct a data protection impact assessment (DPIA)*".

Remove poorly documented concepts. BA and SDE pointed out that some guidelines were too abstract to be implemented. After analyzing the above guidelines and finding that the current literature did not provide enough information to make the guideline actionable, we decided to remove this category of guidelines. For example, in the Fairness component, we removed the "*construct validity*" concept because it is poorly documented and rarely mentioned in the current literature; additionally, there are no specific indications of what mitigations exist and how they can actually have a practical impact through the SDLC.

### 5.3 Lessons learnt

In addition to the improvements discussed in Section 5.2, thanks to the case study, we identified additional insights that may be useful to further improve POLARIS framework and its usability.

In this section, we describe some lessons learnt from the analysis of the case study data.

User Experience (UX). From the testing of POLARIS what clearly emerged is that an Excel sheet provides a suboptimal user experience, and the resulting complexity of use can lead to new users abandoning POLARIS. For this reason, we plan to provide a better UI to make POLARIS consultation easier and faster.

To achieve this, we are evaluating various types of UI: a questionnaire-like web page, a Single Page Application (SPA) — like the VIS-Prise tool proposed in [38] —, or a book-like web page — like the PAI's Guidance for Safe Foundation Model Deployment[7]. We plan to realize one digital prototype for each approach and conduct user acceptance tests, in order to understand which one is the easiest for the users.

Lack of concrete guidelines for the Monitoring phase. While observing the IT user utilizing POLARIS, we noticed that the monitoring SDLC phase is the one with the least concrete guidelines, probably due to the fact that the advent of the cloud has, in some way, delegated the monitoring burden to third parties. Anyway, since the IT user had the necessity to deploy an AI-enabled system on an on-premise infrastructure, he pointed us to this lack of the framework. For this reason, we plan to conduct further research in the current literature to collect as many concrete monitoring guidelines as possible; this way, POLARIS would become more effective in the monitoring phase.

Lack of specific tools and guidelines for LLM. Most of the knowledge sources that compose POLARIS were published before the recent great improvements in Large Language Models (LLMs). As a consequence, POLARIS now lacks guidelines and tools specific to LLMs. As all team members pointed out, LLMs are becoming increasingly used in AI-enabled systems and they plan to use them in in the next version of the application on which we tested POLARIS. For this reason, we plan to research, integrate and experiment with both guidelines and automatic tools to specifically address trustworthiness in LLMs.

---

[6]https://www.enisa.europa.eu/risk-level-tool/risk

[7]https://partnershiponai.org/modeldeployment/

Monitor literature advancements. POLARIS for its nature is a live framework that needs to be regularly updated as the literature in the AI-enabled systems field is constantly evolving. We continuously monitor new grey- and white-literature sources. This way, we can timely discover new knowledge sources (both guidelines and tools) that can be integrated into POLARIS and keep it updated. For example, we are evaluating the new MITRE Atlas mitigations collection[8].

Make the framework open source. The reflections on the previous point lead us to the idea that POLARIS frame- work should be published on an open-source repository, freely accessible under the CC-BY 4.0 license, so that anyone interested in the project can integrate new knowledge or refine the existing one. This way, we could exploit the wisdom of the crowd to further enhance POLARIS. For this reason, we published POLARIS in a public GitHub repository [39].

## 6 LIMITATIONS

In this section, we discuss the limitations of POLARIS framework.

*Trustworhty AI principles*. In this first release of the framework, we focus only on four principles: Explainability, Fairness, Security, and Privacy. We know there exist further TAI principles than those we considered and all of them require attention. However, in this first experiment, we choose to select only the most recurrent principles according to the current literature [10]. The reason was the need to have a shorter development phase of the framework in order to pro- ceed with a first validation in an industrial case study. If the next validations on further case studies will confirm us the stakeholders approve POLARIS and would use it in their daily work, we will continue its development by introducing more TAI principles in future versions.

*Validation*. We know POLARIS must be validated in several industry case studies to further improve it and that a single validation is not enough. However, we plan to do further validations with AI practitioners once we develop a better UI. Having a better usability experience will allow us to validate POLARIS on a growing number of projects with lower effort. This will facilitate the testing and validation step provided by other research groups and companies.

## 7 CONCLUSION

In this work, we described POLARIS, the framework we designed to fill the gaps highlighted in the review of the state of the practice and to provide AI practitioners with actionable guidelines specific to each phase of the SDLC.

POLARIS has four pillars (or *components*), which are Explainability, Fairness, Security, and Privacy. These principles have been chosen as they are the most recurrent TAI principles found in the current literature [10]. Each component provides practical guidelines and tools to support different kinds of stakeholders across the entire SDLC.

Its added value is that it provides knowledge already freely accessible online but in an organized and systematized way.

---
[8]https://atlas.mitre.org/mitigations

We applied POLARIS to an industrial case study, and collected feedback by means of the thinking-aloud method. This allowed us to identify several improvements. Among the most important changes, we improved (i) its usability (e.g., by providing it with filtering capabilities), (ii) removed comparable guidelines, which could lead to confusion or redundancies and (iii) added references to tools when these were not present.

From a usability point of view, we are planning a set of improvements on which we will continue working on. By providing an even more usable UI, we believe we will be able to validate it on a growing number of case studies, thanks to the collaboration with other companies and research groups.

Moreover, if further validations confirm us the stakeholders are interested and plan to use POLARIS, in the next versions we plan to integrate more TAI principles.

POLARIS is a preliminary attempt to organize and make knowledge on TAI principles easily accessible and available to different kinds of stakeholders. It is a pioneering prototype whose goal is to make AI professionals, policymakers, and stakeholders able to navigate the ethical dimensions of TAI with confidence, ensuring that the vast potential of AI is harnessed responsibly for the benefit of society.

## ACKNOWLEDGMENTS

This work was partially supported by project SERICS (PE00000014) under the MUR National Recovery and Resilience Plan funded by the European Union - NextGenerationEU.

## REFERENCES

[1] A. Esteva, B. Kuprel, R. A. Novoa, J. M. Ko, S. M. Swetter, H. M. Blau, S. Thrun, Dermatologist-level classification of skin cancer with deep neural networks, Nature 542 (2017) 115–118.
[2] G. Cornacchia, F. Narducci, A. Ragone, Improving the user experience and the trustworthiness of financial services, in: Human-Computer Interaction - INTERACT 2021 - 18th IFIP TC 13 International Conference, volume 12936 of *Lecture Notes in Computer Science*, Springer, 2021, pp. 264–269.
[3] High-Level Expert Group on AI (AIHLEG), Ethics guidelines for trustworthy AI | Shaping Europe's digital future, 2018.
[4] M. T. Baldassarre, D. Caivano, B. Fernandez Nieto, D. Gigante, A. Ragone, The social impact of generative ai: An analysis on chatgpt, in: Proceedings of the 2023 ACM Conference on Information Technology for Social Good, GoodIT '23, Association for Computing Machinery, New York, NY, USA, 2023, p. 363–373.
[5] European Union, AI Act, 2023.
[6] UNI Global Union, Top 10 Principles for Ethical AI, 2019.
[7] The Public Voice, Universal Guidelines for Artificial Intelligence, 2019.
[8] Google, Tools & Platforms, 2019.
[9] NIST, AI Risk Management Framework, 2019.
[10] A. Jobin, M. Ienca, E. Vayena, The global landscape of ai ethics guidelines, Nature Machine Intelligence 1 (2019) 389–399.
[11] High-Level Expert Group on AI (AIHLEG), High-Level Expert Group on AI (AIHLEG), 2018.
[12] UNESCO Ad Hoc Expert Group (AHEG), Recommendation on the Ethics of Artificial Intelligence, 2021.
[13] Holborn Law LLC, Advisory Council on the Ethical Use of Artificial Intelligence and Data in Singapore, 2018.
[14] NASA Artificial Intelligence Group, NASA Artificial Intelligence Group, 2018.
[15] UK AI Council, UK Government, 2019.
[16] D. Greene, A. L. Hoffmann, L. Stark, Better, nicer, clearer, fairer: A critical assessment of the movement for ethical artificial intelligence and machine learning, in: Hawaii International Conference on System Sciences.

[17] Wagner, B. in *Being Profiled: Cogitas Ergo Sum. 10 Years of 'Profiling the European Citizen'* (eds Bayamlioglu, E., Baraliuc, I., Janssens, L. A. W. & Hildebrandt, M.), Amsterdam University Press, 2018.

[18] L. Floridi, J. Cowls, A unified framework of five principles for ai in society, Issue 1 (2019).

[19] V. S. Barletta, D. Caivano, D. Gigante, A. Ragone, A rapid review of responsible ai frameworks: How to guide the development of ethical ai, in: Proceedings of the 27th International Conference on Evaluation and Assessment in Software Engineering, EASE '23, Association for Computing Machinery, New York, NY, USA, 2023, p. 358–367.

[20] M. T. Baldassarre, D. Gigante, M. Kalinowski, A. Ragone, Survey link, 2023. Https://forms.office.com/e/GVeeWf1Pqz.

[21] F. Zhengxin, Y. Yi, Z. Jingyu, L. Yue, M. Yuechen, L. Qinghua, X. Xiwei, W. Jeff, W. Chen, Z. Shuai, C. Shiping, Mlops spanning whole machine learning life cycle: A survey, ArXiv abs/2304.07296 (2023).

[22] W. Jin, J. Fan, D. Gromala, P. Pasquier, G. Hamarneh, Euca: the end-user-centered explainable ai framework (2021).

[23] S. Muhammad, The fairness handbook, 2022.

[24] ENISA, Securing machine learning algorithms, 2021.

[25] Tensorflow, Responsible ai in your ml workflow (2021).

[26] D. CSIRO, Responsible ai pattern catalogue (2022).

[27] ICO, Guidance on AI and data protection, 2021.

[28] Microsoft, Threat modeling AI/ML systems and dependencies, 2021.

[29] M. T. Baldassarre, D. Gigante, M. Kalinowski, A. Ragone, Polaris appendix, 2023. Https://figshare.com/s/16bc31211b11c3155f81.

[30] A. Wildberger, Alleviating the opacity of neural networks, in: Proceedings of 1994 IEEE International Conference on Neural Networks (ICNN'94), volume 4, pp. 2373–2376 vol.4.

[31] S. M. Lundberg, G. Erion, H. Chen, A. DeGrave, J. M. Prutkin, B. Nair, R. Katz, J. Himmelfarb, N. Bansal, S.-I. Lee, From local explanations to global understanding with explainable ai for trees, Nature Machine Intelligence 2 (2020) 56–67.

[32] P. Runeson, M. Höst, Guidelines for conducting and reporting case study research in software engineering, Empirical Software Engineering 14 (2009) 131–164.

[33] B. Jahne, Computer vision and applications: a guide for students and practitioners, Elsevier, 2000.

[34] C. Lewis, Using the" thinking-aloud" method in cognitive interface design, IBM TJ Watson Research Center Yorktown Heights, NY, 1982.

[35] H. Kuusela, P. Paul, A comparison of concurrent and retrospective verbal protocol analysis, The American Journal of Psychology113 (2000) 387–404.

[36] J. W. Creswell, Research design: Qualitative, quantitative, and mixed methods approaches, Sage, 2013.

[37] M. L. Mohus, J. Li, Adversarial robustness in unsupervised machine learning: A systematic review, 2023.

[38] M. T. Baldassarre, V. S. Barletta, G. Dimauro, D. Gigante, A. Pagano, A. Piccinno, Supporting secure agile development: The vis-prise tool, in: Proceedings of the 2022 International Conference on Advanced Visual Interfaces, AVI 2022, Association for Computing Machinery, New York, NY, USA, 2022.

[39] M. T. Baldassarre, D. Gigante, M. Kalinowski, A. Ragone, Polaris github link, 2023. Https://github.com/dom976/POLARIS_framework.